\documentclass[12pt,preprint]{aastex}

\usepackage{graphicx}
\usepackage{color}
\usepackage{amsmath}
\usepackage{natbib}

\newcommand{\kms}{km~s$^{-1}$}

\shorttitle{NGC 1275 at z=0.7?}
\shortauthors{Heckman et~al.}

\begin{document}

\title{SDSSJ092712.65+294344.0: NGC 1275 at z=0.7?}
\author{Timothy M. Heckman\altaffilmark{1}, Julian H. Krolik\altaffilmark{2},
Sean M. Moran\altaffilmark{3}, Jeremy Schnittman\altaffilmark{4},
Suvi Gezari\altaffilmark{5}}
\altaffiltext{1}{Johns Hopkins University; {\tt heckman@pha.jhu.edu}}
\altaffiltext{2}{Johns Hopkins University; {\tt jhk@jhu.edu}}
\altaffiltext{3}{Johns Hopkins University; {\tt moran@pha.jhu.edu}}
\altaffiltext{4}{Johns Hopkins University; {\tt schnittm@pha.jhu.edu}}
\altaffiltext{5}{Johns Hopkins University; {\tt suvi@pha.jhu.edu}}

\begin{abstract}
SDSSJ092712.65+294344.0 was identified by the SDSS as a quasar, but has
the unusual property of having two emission line systems offset by 2650~\kms.
One of these contains the usual combination of broad and narrow lines, the other
only narrow lines.  In the first paper commenting on this system \citep{kom08},
it was interpreted as a galaxy in which a pair of black holes had merged,
imparting a several thousand km/s recoil to the new, larger black hole.  In
two other papers \citep{bogdan08,dotti08}, it was interpreted as a small-separation
binary black hole.   We propose a new interpretation: that this system is a more
distant analog of NGC~1275, a large and a small galaxy interacting near the center
of a rich cluster.
\end{abstract}


\maketitle

\section{Introduction}
\label{intro}
SDSSJ092712.65+294344.0 (henceforward in this paper ``SDSSJ0927'')
is an object whose spectrum was acquired by
the Sloan Digital Sky Survey (SDSS) as part of its quasar sample.
It was first given special attention by \citet{kom08}, who combed
the SDSS quasar database for objects in which there was a substantial
velocity offset between the [OIII]5007 line and the other emission
lines.   In this case, there is [OIII]5007 emission at the redshift of
the broad line centroid ($z = 0.698\pm 0.001$). Hereafter we will refer to this as
the associated system. There is a second [OIII]5007 line shifted 2650~\kms\ 
to the red, at $z=0.713$ (hereafter the redshifted system). In fact, there
are entire sets of narrow emission lines ([OIII]5007, H$\beta$, [NeIII]3869,
[OII]3727, [NeV]3426) at both redshifts, while broad lines (H$\beta$,
H$\gamma$, MgII2800) are present only in the lower redshift associated
system.  Moreover, the narrow lines
in the associated system all have substantially broader profiles
(FWHM $\simeq 450$--2000~\kms) than do those in the
redshifted system (FWHM $\simeq 170$~\kms).  No stellar lines can
be detected in the SDSS spectrum, so there is as yet no measure
of the redshift of the host galaxy. For reference, the SDSS spectrum is shown in
Figure~1 with the key lines from both systems marked.

\begin{figure}
\centering
\includegraphics[width=0.75\columnwidth]{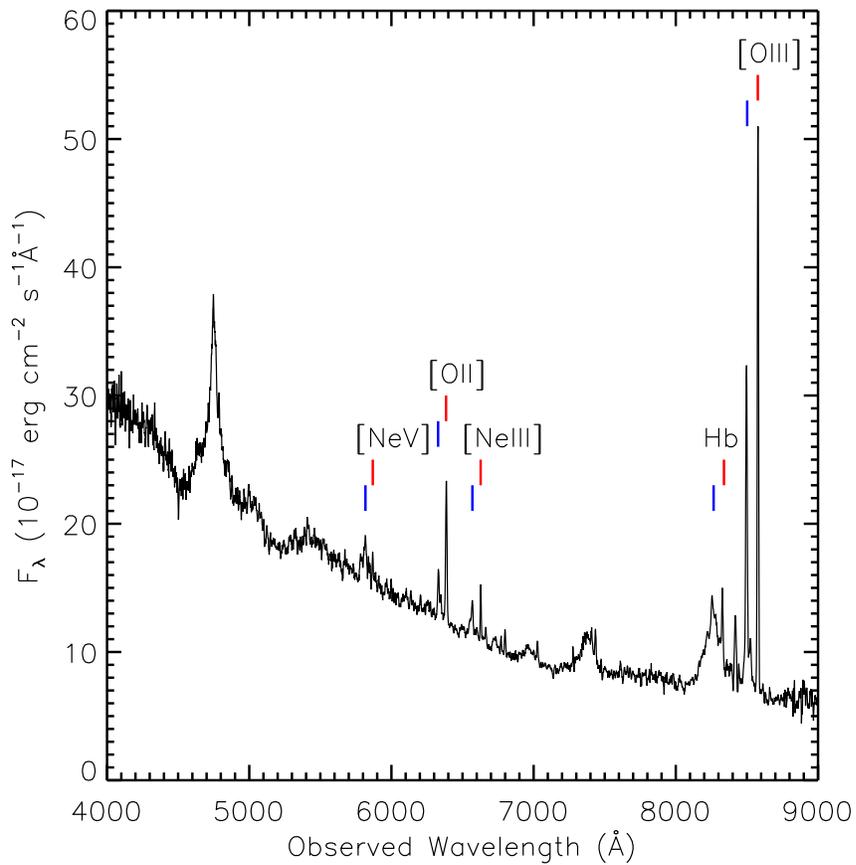}
\caption{SDSS spectrum of SDSSJ0927 with the key narrow emission lines from 
the associated system ($z=0.698$; lower, blue vertical ticks) and redshifted system ($z=0.713$; upper, red ticks) marked. From high to low wavelength, broad H$\beta$, H$\gamma$, H$\delta$, and MgII can also be seen.}

\end{figure}

Recent numerical relativity calculations have shown that it is possible
for a black hole merger to be sufficiently asymmetric that the new
merged system moves relative to the original center of mass by as
much as several thousand km/s when the spins of the original pair
are properly aligned relative to the orbital angular momentum and each other
\citep{baker06,camp2007,gonz07,pollney07}.  On this basis, \citet{kom08}
suggested that this is an example of exactly such an event: the redshifted
narrow emission lines indicate the rest-frame of the host galaxy, while
the associated emission lines, both broad and narrow, are attached to the
black hole of the ejected quasar, which is moving through the host with a
velocity along our line-of-sight of 2650~\kms.

This suggestion was criticized by \citet{bogdan08} and \citet{dotti08},
who both argued that the data were more plausibly explained as arising
from a black hole binary of separation $\sim 0.1$--0.3~pc and mass
ratio $q \simeq 0.1$--0.3.  In this picture, both black holes are
surrounded by a circumbinary disk whose inner edge is a few times
the orbital separation.  The less massive black hole captures the
majority of any gas travelling inward from the inner edge of the disk;
accretion of this gas powers the quasar. Thus, the associated system
would be attached to the lower-mass black hole, while the redshifted system
would be affiliated with the more massive black hole. Both \citet{dotti08} and
\citet{bogdan08} argue that this model is much more probable than the
merger-recoil model.  However, \citet{dotti08} estimate that the rate of
production of such systems should be $\sim 10^{-6}$~yr$^{-1}$ within
$z \simeq 1$, while \citet{bogdan08} argue that such a recoiling quasar
would stay within the galaxy for $\sim 10^7$~yr; in other words, there
should be $\sim 10$ at any given time within $z \simeq 1$.   On the
other hand, there is no guarantee that the merger product would
carry enough accretion fuel to last for $10^7$~yr.  In addition,
it seems somewhat implausible that
a recoiling black hole would create a narrow emission line system
centered on its velocity.  The gas emitting these lines
would need to have had initial binding energy to the pre-merger black
hole within $(v_{NLR}/v_{\rm kick})^2 \simeq 3\%$ of the critical binding
energy at which matter stays bound to the merged black hole.  Here
$v_{NLR}$ is the characteristic speed of the gas relative to the
merger remnant and $v_{\rm kick}$ is the recoil speed.

Serious difficulties also exist for the binary model.  The
problem we find most disturbing is the tight kinematic relation between
the broad emission line system and its associated narrow lines.  In
this model, the broad lines are made in more or less conventional
fashion, even though the second, more massive black hole, is close
enough that its gravity must influence the broad line gas---after all,
its gravity is strong enough to force an orbital speed for the quasar
black hole $\geq 2650$~\kms. The associated narrow lines
are attributed to gas flowing between the circumbinary disk and the
lower-mass black hole, and therefore traverse a region where, by
construction, the characteristic orbital speed is $\gtrsim 2000$~\kms.
It is hard to understand how, under these circumstances, the narrow lines
could be centered on the velocity of the broad lines and
have FWHMs as small as 450~\kms.  In addition, \citet{dotti08} estimated
that the density of the associated narrow emission line gas, if it
occupies a thin disk filling the area within the edge of the circumbinary
disk, is $\simeq 2$--8$\times 10^6$~cm$^{-3}$.  Because it could be
clumpy, this is only a lower bound (in fact, \cite{bogdan08} suggest
that this gas could be the narrow accretion stream linking the inner
edge of the circumbinary disk and the accretion disk of the smaller
black hole; if so, there would perforce be very strong clumping).
However, at densities this high, the [OIII]5007 line is
at least partially suppressed by collisional de-excitation, and
the [OII]3727 line, whose critical density is $\sim 10^4$~cm$^{-3}$,
is even more so. 

In view of these issues, we propose an entirely different model for
this system, which we believe fits the existing data well and is
intrinsically more plausible: that this system is a high-redshift
analog to the familiar nearby Seyfert galaxy NGC~1275.  This galaxy
(also known as 3C~84 and Perseus~A) is located at the center of the
Perseus cluster of galaxies.  Its optical emission line spectrum, judged
by the standard of anything other than SDSSJ0927, is
most peculiar: it has two distinct narrow emission line systems,
one close to the redshift of the galaxy, the other 3000~\kms\  to
the red \citep{mink57}.  Emission line imaging has shown that the redshifted
system is displaced from the center of the galaxy by 10--20~kpc \citep{consel01},
while 21~cm and X-ray absorption at the higher redshift demonstrates that
this gas lies between us and the nucleus of NGC~1275 \citep{deY73}.
Although argument raged for decades over whether the redshifted system
is part of a small galaxy interacting with NGC~1275 (as originally
proposed by \citet[]{mink57}) or is a member of the Perseus cluster
that is merely superposed on NGC~1275 \citep[e.g.,][]{rubin77}, the
weight of the evidence now rests on the side of interaction \citep{consel01}.
To acquire its velocity, the smaller galaxy presumably has fallen
from the outskirts of the Perseus cluster on a nearly radial orbit;
given the cluster dispersion of 1300~\kms  \citep{struble91}, such
a supposition is quantitatively reasonable. 
\footnote{Since submitting this paper, a paper by Shields et al. (2008) has appeared that presents new data and independent arguments that SDSSJ0927 is most likely to be a superposition of two systems rather than an ejected or binary black hole.}

Almost all of these properties are duplicated in SDSSJ0927.
Indeed, the only difference is that the nucleus of NGC 1275
is a moderately luminous Type 2 Seyfert, while SDSSJ0927 contains a powerful quasar.
Moreover, all the difficulties of the models proposed earlier (either
the high-speed recoil or the binary black hole model) disappear
viewed from this perspective:  There is no problem with probabilities---we
already know of another example in the nearby Universe.  There is
no problem with keeping the gas of the higher redshift system within
a narrow range of velocities---it is confined within a galaxy while
it travels through a region (the cluster now, not the vicinity of
a binary black hole) with much higher orbital speeds.  In addition,
as we will show shortly, the directly derived properties of the emission-line gas
are compatible with this model.

\section{Emission Line Analysis}

The flux ratios of the associated narrow lines
are very similar to those commonly found in AGN.  According to
\citet{kom08}, the lines H$\beta$, [OII]3727, [OIII]5007, [NeIII]3869,
and [NeV]3426 have the relative strengths 1.0, 1.5, 6.7, 1.8, and 4.0 (Table~1),
while in the ``mean'' narrow line system in a Seyfert or quasar, they would be 1.0, 3.2, 5.0,1.4, and 1.2 \citep{kro99}.  The greatest contrast is in the [NeV]3426
line.  We might therefore suppose that the physical conditions in this
gas (ionization parameter, density, etc.) are very similar to what is
generally found in other AGN.  In our new interpretation, this gas
can therefore be found in its usual location, $\sim 100$---1000~pc
from the black hole at the center of the galaxy.

\begin{deluxetable}{lccc}
\tablecolumns{4}
\small
\tablewidth{0pt}
\tablecaption{Relative Line Strengths}
\tablehead{\colhead{Line} & \colhead{Associated System} & \colhead{Redshifted System} & \colhead{Mean Narrow Line System}}
\startdata
H$\beta$\tablenotemark{a} & 1.0 & 1.0 & 1.0 \\
$\textrm{[OII]}3727$ & 1.5 & 2.6 & 3.2 \\
$\textrm{[OIII]}5007$ & 6.7 & 10.1 & 5.0 \\
$\textrm{[NeIII]}3869$ & 1.8 & 0.7 & 1.4 \\
$\textrm{[NeV]}3426$ & 4.0 & 0.3 & 1.2 \\

\enddata
\tablenotetext{a}{Normalized to H$\beta$}
\end{deluxetable}

The detailed character of this emission line system is one place
where the analogy with NGC~1275 may break down. The line
emission at the systemic velocity in NGC~1275 is likely affected by a
number of processes in addition to AGN photoionization
\citep{johnst88,sabra00,ferl08} that may have their origin in special properties
of the cluster environment.   Some of these processes may also influence the
narrow emission lines associated directly with the quasar SDSSJ0927.
However, its luminosity is so much greater than that of the Seyfert
galaxy NGC~1275 that the relative importance of these other processes
may be smaller; perhaps that is why its line ratios so closely
resemble those of a generic AGN.

The relative fluxes of the narrow lines in the redshifted system in SDSSJ0927 are 
also similar to the narrow emission-line regions
in AGN: 1.0, 2.6, 10.1, 0.7, and 0.3, where the lines are listed in the same
order as before and also given in Table~1.  In the model we favor, this emission would come from gas in
the small galaxy that has fallen in from outer regions of the cluster, but is
now close enough to the quasar that the gas is photoionized by the quasar
continuum.  The power source for the redshifted emission lines is different
from the case of NGC~1275 where 
(owing to the relatively small ionizing luminosity of the AGN) the emitting gas
in the infalling galaxy is photoionized by hot stars within the galaxy itself. 

Without going into an
extensive photoionization analysis, we can use the emission-lines in the redshifted
system in SDSSJ0927 to see if its basic physical properties are consistent with
our model. The flux ratio of the two members of the [OII]3726,3729 doublet is
1$\pm$0.1, implying an electron density $n_e \sim300$ cm$^{-3}$ \citep{of06}.
In our model this would be gas that, in the absence of a nearby quasar,
would reside in cold HI clouds in the ISM of a normal galaxy. The
emission measure inferred from the H$\beta$ luminosity is
$1 \times 10^{67}$~cm$^3$ (we remeasured both the H$\beta$ and the continuum flux
in the SDSS spectrum, and assume $H_0 = 70$~\kms~Mpc$^{-1}$, $\Omega_M=0.3$,
$\Omega_\Lambda=0.7$). The implied gas mass is then simply the
emission measure divided by the mean density, so that
$M_g \sim 2.5 \times 10^7 M_{\odot}$, an entirely plausible
gas mass for the irradiated dense clouds in the hypothesized infalling galaxy.

We can estimate the distance from the QSO to this infalling galaxy by using
the emission line ratios to estimate the ionization parameter in the gas due
to exposure to the quasar continuum:
$r \sim [L_{ion}/(n_e\xi)]^{1/2}$, where $L_{ion}$ is the ionizing luminosity
of the quasar, $r$ is the distance from the gas to the quasar, and
$\xi\equiv L_{ion}/(nr^2)$ is the ionization parameter.

Assuming that $F_{ion}/[\lambda F_\lambda (4000\AA)] = 5$, we find that
the quasar ionizing luminosity is $8 \times 10^{45}$~erg~s$^{-1}$. 
The ionization parameter appropriate to the observed [OIII]5007/H$\beta$ ratio is
$\xi \sim 0.05$ \citep{of06}, while we measure $n_e = 300$ cm$^{-3}$
(we caution, however, that the [NeV] line indicates that there is
a significant amount of gas in a higher ionization state, so a simple
one-zone model may be somewhat misleading).
The implied physical separation between the quasar and hypothesized infalling
galaxy is then 8~kpc
(similar to the transverse separation between the two systems seen in NGC~1275). 
If the typical column density of this gas is comparable to the column density of
gas in a galactic disk, $N_H \sim 10^{21}$~cm$^{-2}$, it would be
optically thick at the Lyman edge, so all the incident ionizing photons
would be absorbed and reprocessed. The ratio of the ionizing luminosity
from the quasar to the narrow H$\beta$ emission line implies that the irradiated gas
would cover a solid angle of $\sim$0.2 steradians, or an area of 12 kpc$^2$
for a separation of 8 kpc.

Thus, in all respects that we can measure from the existing data, the properties
of the redshifted narrow emission-line system are fully compatible with our
hypothesis that SDSSJ0927 is a higher redshift version of the NGC~1275 system:
a galaxy falling into the deep potential well of a rich cluster of galaxies
where it interacts with the host galaxy of an AGN.

\section{Testable predictions}

This model makes a number of predictions that can be readily tested by
observations.  First, if stellar absorption lines can be detected in
this object's spectrum, their centroid should be close to the bluer system's
velocity, and they might be doubled, with a separation $\simeq 2600$~\kms .
We recognize, however, that it is in general difficult to detect stellar
absorption lines in quasar spectra because they are usually so strongly
diluted by the quasar continuum. 
In addition, high-resolution optical imaging (e.g., using {\it HST}),
should detect some extended light, perhaps exhibiting irregular structure
due to the galaxy-galaxy interaction we suggest is occurring.

Secondly, there should be a rich cluster surrounding
this object.  The left panel of Figure~2 shows a 
color SDSS cutout 2\arcmin\ on a side, centered on SDSSJ0927. 
This corresponds to a physical size of $\sim850\times 850$~kpc at the 
redshift of the quasar.  A number of faint red galaxies are visible,
particularly to the southwest of the quasar.  In the right-hand panel, we plot a 
histogram of photometric redshifts for galaxies in this field, using redshifts
derived from template fitting and made available in the
{\tt Photoz} table \citep{csabai03}.  Typical quoted errors for the photometric
redshifts are $\pm 0.05$.  It is clear that in the region
as a whole there is an overdensity of galaxies with redshifts slightly
less than that of the quasar; the small shift is likely an artifact of
the photo-$z$ template fitting, rather than a real difference between the
quasar redshift and that of the galaxies in this concentration.  
Particularly noteworthy, nearly all the faint red galaxies noted in the SW
quadrant of the image have redshifts within 2 standard errors of 0.7 
(shaded part of the histogram); the two exceptions in the SW quadrant with
photo-$z$s of $z\sim0.45$ also have photo-$z$s
determined though the alternative, neural network method, and these estimates
would place them at $z \simeq 0.65$ as well.  Each
of these galaxies, if actually at $z \simeq 0.7$, would be within 425~kpc
(projected) of SDSSJ0927. 

Although the existence of $\sim10$ co-located galaxies does not ensure
that a massive cluster is present, at least five of the red galaxies in
the SW quadrant have $z$-band absolute magnitudes
in the range $-22 < M_z < -24$, again assuming that they are at $z=0.7$.  
At this luminosity, each of these galaxies would have a stellar
mass $\simeq 10^{11\pm 0.5} M_\odot$ if they have SEDs
typical of early-type cluster galaxies at $z=0.7$.  Such a concentration of massive
galaxies within a few hundred kpc of each other suggests that we are indeed
viewing the brightest members of a massive cluster, and not just a smaller
group of galaxies.

Further evidence in favor of both a galaxy cluster surrounding this quasar
and interaction between the quasar's host galaxy and members of this cluster
comes from the deep $\chi^2$ co-addition of all five SDSS images shown 
in Figure~3.  This image
reveals that SDSSJ0927 has two close companions within 40~kpc projected
radius, once again assuming that they, too, are at $z=0.7$. The object
to the lower left (SE) has an entry in the SDSS photometric catalog and a
photo-$z$ placing it in the apparent cluster at $z=0.6$--0.7.  The object
to the north of the quasar has no redshift
estimate, but its larger spatial extent stretching toward the quasar
suggests the possibility that light from this galaxy may contaminate the
SDSS spectrum.

\begin{figure}\label{fig:cluster}
\includegraphics[width=0.5\columnwidth]{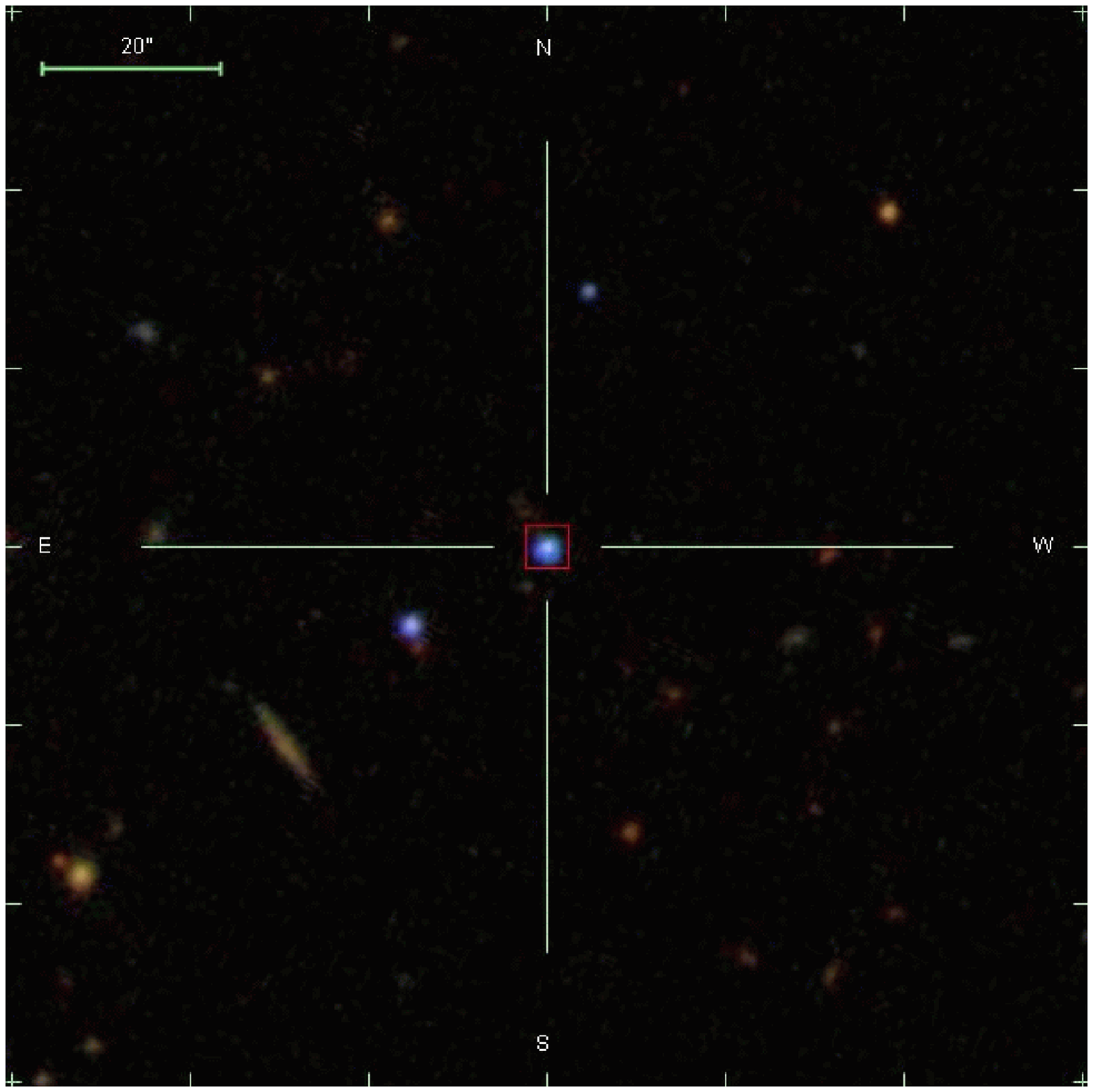}
\includegraphics[width=0.5\columnwidth]{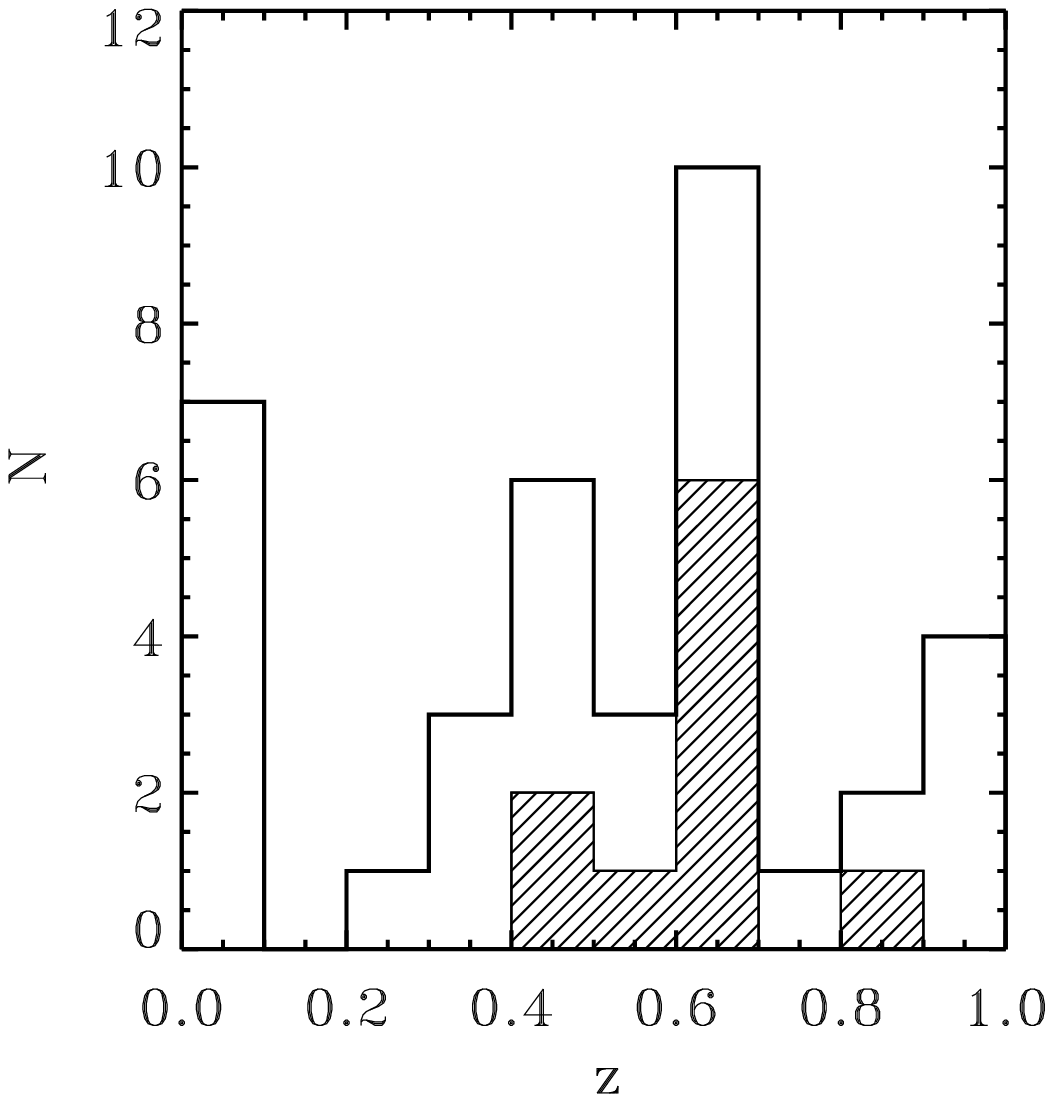}
\caption{Left: SDSS color cutout of a $2\arcmin\times 2\arcmin$ box around
SDSSJ0927. The quasar is marked with a red box. 
Right: Histogram of SDSS-derived photometric redshifts for all galaxies in
the left panel within 1\arcmin\ radius of the quasar.  The redshifts of the
faint red galaxies to the southwest of the quasar are shown by shading in
the histogram.}
\end{figure}

\begin{figure}\label{fig:chisq}
\centering
\includegraphics[width=0.75\columnwidth]{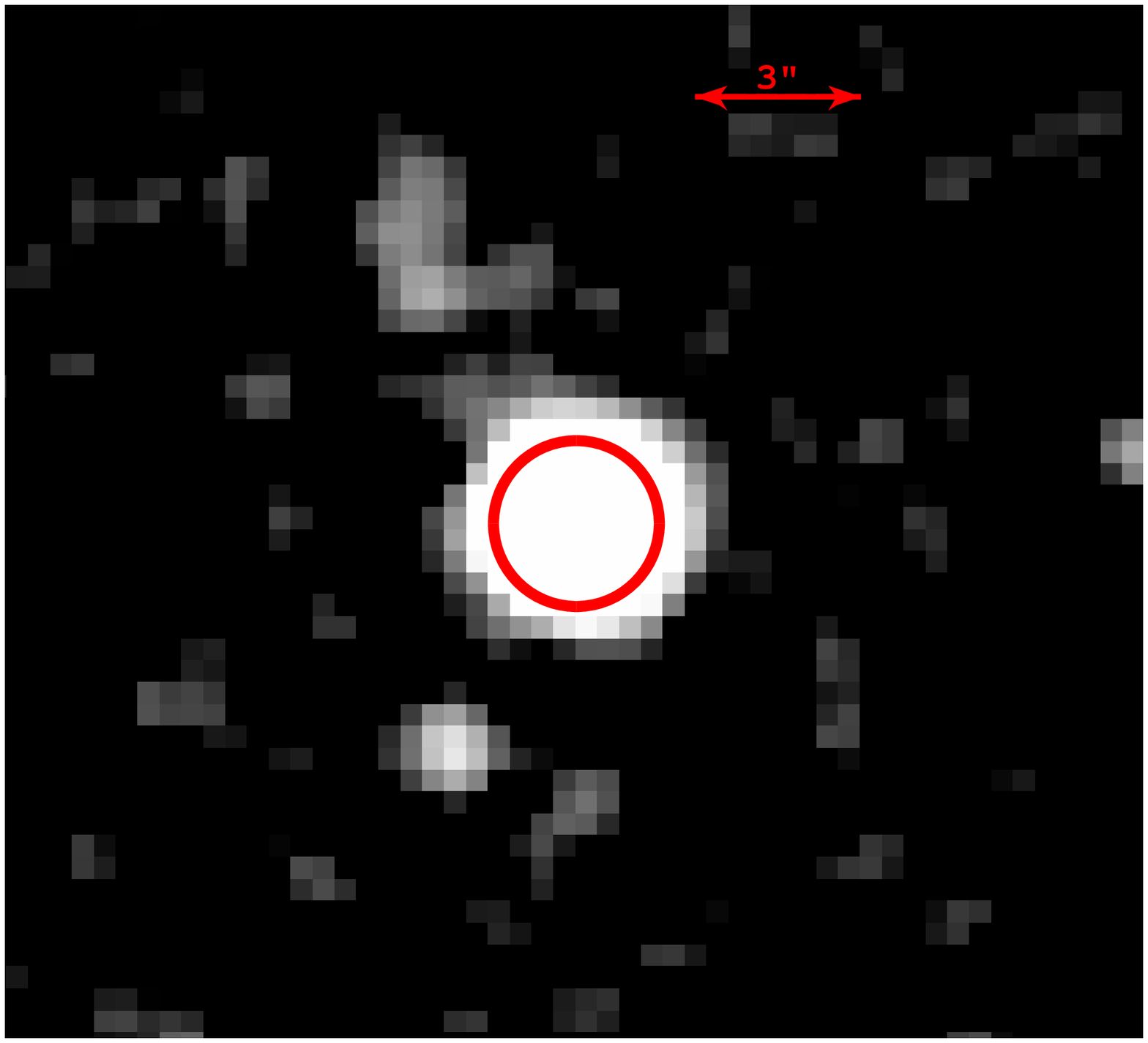}
\caption{Postage stamp $\chi^2$ image of the immediate environment
around SDSSJ0927 produced by combining all five SDSS images ({\it ugriz}
bands), and then smoothing with a Gaussian of radius 2 pixels (1.8" FWHM).
All non-black pixels represent a significant detection of object flux, and
the red circle indicates the size of the SDSS spectroscopic fiber.}
\end{figure}

A third prediction is that spatially-resolved spectroscopy should
reveal a spatial offset between the two narrow line systems.  At
the distance of SDSSJ0927, 8~kpc corresponds to $\simeq 1.1^{\prime\prime}$.
Even if the system is partially projected along the line of sight, 
for most orientations the expected separation should be large enough for a
long-slit spectrum to reveal easily any shift in centroid between the two
emission line sets.  In fact, the NE galaxy seen in Figure~3
actually extends to very close to the position of the quasar.
Perhaps the redshifted emission-line system is associated with the portion of
this galaxy nearest to the quasar (inside the SDSS $3^{\prime\prime}$ fiber). 

Fourth, a cluster whose potential is deep enough to produce an orbital
speed of 2600~\kms\ should be a bright X-ray source.  \citet{kom08}
found two observations in the Rosat archives and estimated the X-ray
luminosity at $5 \times 10^{44}$~erg~s$^{-1}$.  Although an X-ray luminosity
of this magnitude would be expected from a quasar having the observed
optical luminosity, this luminosity would also be typical of rich clusters
of galaxies.  It is possible that the X-rays come from both sources.

If these observations had been
pointed toward SDSSJ0927, the angular resolution of Rosat (half-power
width $\simeq 5^{\prime\prime}$) would have been great enough to
clearly distinguish a point-source from an extended source, as this
half-power width translates to 36~kpc at $z = 0.7$.  However, in both
observations, SDSSJ0927 was near
the edge of the field of view, so the resolution was somewhat
degraded by imperfections in the Rosat mirror.  In addition, confidence in
an observational estimate of this source's angular extent is also undercut
by the small number of counts detected: 70 in one observation, 47 in the other.
Neither {\it Chandra} nor {\it XMM-Newton} observations of this field have
been obtained so far.
  
\section{Conclusions}

    We believe that, in contrast to more exotic interpretations involving
binary black holes and black hole merger events, the most plausible
interpretation of the SDSS quasar SDSSJ0927 is that it
is a system much like NGC~1275: an AGN lying near the center of a rich
cluster of galaxies, interacting with a smaller galaxy that has fallen
toward it from farther out in the cluster.  There is a very strong
phenomenological resemblance between the two systems (the offset
velocity is actually about $10\%$ smaller in the SDSS object), and
there are none of the physical conundra that cause concern about the
black hole binary interpretations.

     Our arguments also suggest that the correct interpretation of
double-peaked emission line profiles can be a subtle problem.  Two
well-separated peaks in a profile do demand two separate regions with
dynamical coherence in which gas suitable for line emission is gathered.
How that coherence is maintained is another matter: a pair of
massive black holes could account for two deep gravitational potential
minima, but so could two dense stellar clusters, or one black hole
and one stellar cluster.  For that matter, there
can also be cases in which the gas concentrations are not due to
gravity, but to jet working surfaces.   Moreover, as we have pointed out,
two separate line emission regions do not necessarily require two
separate sources of ionizing photons.  The strongest evidence for
two sources of ionizing photons is two spatially separated regions
of continuum emission.

\vspace{0.35cm}\noindent This work was partially supported by
the NASA LTSA program (TMH) and NSF grant AST-0507455 (JHK).  We would also
like to thank the organizers of the JHU Monday evening ``wine
and cheese'' seminar for providing the stimulus for this project.


\end{document}